\documentclass{ws-ijmpe}
\usepackage{enumerate}

\newcommand{\Mpl}{M_{\rm Pl}}
\newcommand{\lpl}{\ell_{\rm Pl}}
\newcommand{\Mf}{M_\star}

\begin{document}

\markboth{M. Bleicher, P. Nicolini, M. Sprenger and E. Winstanley}{Micro black holes in the laboratory}

\catchline{}{}{}{}{}

\title{MICRO BLACK HOLES IN THE LABORATORY}

\author{MARCUS BLEICHER, PIERO NICOLINI, MARTIN SPRENGER}

\address{Frankfurt Institute for Advanced Studies (FIAS), Johann Wolfgang Goethe University, 
Ruth-Moufang-Strasse 1\\ Frankfurt am Main, 60438, Germany\\
Institut f\"ur Theoretische Physik, Johann Wolfgang Goethe-Universit\"at,
Max-von-Laue-Strasse 1\\ Frankfurt am Main, 60438, Germany\\
bleicher@fias.uni-frankfurt.de, nicolini@fias.uni-frankfurt.de,
sprenger@fias.uni-frankfurt.de}

\author{ELIZABETH WINSTANLEY}

\address{School of Mathematics and Statistics, The University of Sheffield,
Hicks Building, Hounsfield Road\\ Sheffield, S3 7RH, United Kingdom\\
E.Winstanley@sheffield.ac.uk}

\maketitle

\begin{history}
\received{(received date)}
\revised{(revised date)}
\end{history}

\begin{abstract}
The possibility of creating microscopic black holes is one of the most exciting predictions for the LHC, with potentially major consequences for our current understanding of physics.
We briefly review the theoretical motivation for micro black hole production, and our understanding of their subsequent evolution.
Recent work on modelling the radiation from quantum-gravity-corrected black holes is also discussed.
\end{abstract}

\section{Introduction}

The possibility of producing microscopic black holes in particle detectors is one of the most intriguing
predictions of  recent  high energy physics. Since the original ideas (developed nearly simultaneously  by Dimopoulos and Landsberg\cite{DiL01} and Giddings and Thomas\cite{Giddings:2001bu}),  the topic has attracted considerable media exposure\footnote{See for instance ``Physicists Strive to Build A Black Hole'', New York Times, September 11, 2001.
} and has generated a huge scientific literature (for one of the latest reviews see Ref. \refcite{Kan08}).
Against this background, there is also increasing skepticism: man-made micro black holes require a modification of standard Einstein gravity, namely the introduction of additional spatial dimensions, that according to recent data are harder to detect than some hoped\cite{FGG11}. In addition, the latest experimental investigations, excluding the formation of black holes with masses up to $4.5$ TeV, are further supporting the highly speculative character of the topic\cite{CMS11}.
As a consequence, we may ask why microscopic black holes are still so important, ten years after their production was conjectured.
The answer lies in the fact that microscopic black holes have the potential to reveal deep insights into fundamental physics, in particular quantum gravity.

Quantum gravity is an ongoing attempt to reconcile quantum mechanics with general relativity, aiming to unify into a single consistent model all known observable interactions in the universe, at both subatomic and astronomical scales. Despite decades of efforts and the formulation of many candidate theories, our understanding of quantum gravity is far from being complete. On the experimental side, we do not yet have observations to confirm or reject theoretical formulations, due to the extreme energy required to observe quantum gravity effects. A possible way out of this puzzling situation is given by a prototype of quantum gravity, namely a model theory that can be used to test the concepts and processes we are following in the path to quantum gravity. This is quantum field theory (QFT) in curved space-time, the simplest theoretical arena for studying  particle physics in the presence of gravitational effects. QFT in curved space, which is often regarded as the semi-classical limit of quantum gravity, has a robust theoretical prediction: a black hole emits quantum thermal radiation like a black body at a temperature proportional to the inverse of its mass, $T\propto 1/M$. This result, known as black hole evaporation since Hawking's early work\cite{Haw75},  tells us that astrophysical black holes have negligible thermal properties, while only smaller size black holes could have temperatures relevant for experiments on Earth. One can show that the black hole evaporation time would exceed the age of the universe for black holes with sizes bigger than $10^{-16}$ m.
Thus microscopic black holes provide the best avenue for the observation of Hawking radiation.
%
%
\section{Black holes at the terascale}

From the aforementioned estimates of evaporating black hole sizes it is evident that we are concerned with a topic in the realm of particle physics. Thus we may ask upon what conditions an elementary particle of mass $M$ can be a black hole. To answer this question we proceed by steps. First we have to think how to shrink a volume containing a mass $M$ as much as possible. When the size of the volume approaches the gravitational radius $r_g\approx GM/c^2$, the system will undergo gravitational collapse to form a black hole and cannot be made smaller. As a second step we have to consider that in the microscopic world quantum mechanical effects cannot be ignored: uncertainty relations imply that we cannot know the position of a particle with better accuracy than $\lambda\approx\hbar/Mc$.\footnote{In our estimates we do not distinguish between $h$ and $\hbar$, thus taking $2\pi\approx 1$. Sometimes these are informally called ``Feynman units''.\cite{Adl10}} Smaller scales would imply an uncertainty in energy greater than $Mc^2$, which is enough energy to create a pair of particles of the same type. The condition for a particle black hole is dictated by the overall minimum, which occurs when the two fundamental length scales, $r_g$ and $\lambda$, equal
$ \hbar/Mc\approx GM/c^2.$
This implies that the formation of particle black holes takes place at the Planck scale, namely $M\approx \Mpl\equiv \sqrt{\hbar c/G}$ and $r_g\approx\lambda\approx\lpl\equiv\sqrt{\hbar G/c^3}$. This is highly problematic since it requires energies far beyond that of any ground based particle physics experiments and even of the highest energy cosmic ray ever detected\cite{Bir93}.

Therefore, to have any hope of producing particle black holes, we must accept the existence of some ``mechanism'' for lowering the gravitational coupling to a new fundamental energy scale $\Mf$ accessible to current or near future experiments. One such mechanism is a modification of gravity at short length scales, by allowing space-time to have $k$ additional spatial dimensions, called extra dimensions. Following Refs. \refcite{AAD98}, the usual four dimensional universe would be a slice, called a brane, inside a $(4+k)$-dimensional space-time, called the bulk. Standard model fields would be constrained on the brane, while only gravity would be allowed to propagate in the bulk. The extra dimensions must be enough small to be usually unobservable and large enough to significantly lower the fundamental scale to accessible values. Thus by defining $R$, the size of each extra dimension, we require Newton's law to be valid at large distances
\begin{equation}
\phi(r)=\frac{\hbar^{k+1}}{M_\star^{2+k}}\frac{c^{1-k}}{R^k}\frac{M}{r} \rightarrow \frac{\hbar c}{\Mpl^{2}}\frac{M}{r}\quad \mathrm{as} \ \ r\gg R.
\label{newton}
\end{equation}
Conversely we want $R$ and $k$ to be such that
\begin{eqnarray}
M_\star \approx \left(\frac{\lpl}{R}\right)^{\frac{k}{k+2}} \Mpl \sim 1\ \mathrm{TeV}.
\end{eqnarray}
Higher dimensional black hole geometries have been known since early studies in string theory.\cite{MyP86} Our $(4+k)$-dimensional space-time no longer possesses the full, higher-dimensional space-time symmetries due to the presence of the brane.  However, if the black hole size is smaller than $R$, the brane tension is negligible and the usual higher-dimensional, spherically symmetric metrics are an acceptable approximation to the actual space-time geometry.
For TeV black holes, i.e. $M\approx 1 \ \mathrm{TeV}$, we find from (\ref{newton}) that $r_g\approx(M/\Mf^{k+2})^{1/(k+1)}\sim 10^{-19}\ \mathrm{m}$. Thus we conclude that the size of each extra dimension must be $R\gg 10^{-19}\ \mathrm{m}$ to have a negligible brane tension.
The latest data from the LHC severely constrains the size of extra dimensions. Current limits are $R\lesssim 10^{-12}\ \mathrm{m}$ which imply $k \gtrsim 3$ to have $\Mf$ at the terascale.\cite{FGG11}
Despite these tight limits on $R$, the black hole is enough small, $r_g\ll R$, but it is also big enough to be plentifully produced in hadronic collisions\cite{Bleicher:2001kh}. A rough upper limit for the black hole production cross section is given by the ``black disk'' estimate $\sigma\approx \pi r_g^2\sim 100 \ \mathrm{pb}$. At the current LHC peak luminosity $L\sim 10^{37}\ \mathrm{m}^{-2}\ \mathrm{s}^{-1}$ roughly one black hole every three seconds would be produced. Even if this is an optimistic estimate since a variety of effects have been ignored in this calculation,\cite{Cav03} we obtain a more realistic value of roughly $\sim 10^2 \ \mathrm{BHs}/\mathrm{year}$, which is what one expects from the claimed production of black holes in cosmic ray showers.\cite{FeS02}

The life of a microscopic black hole in a particle detector is still not fully understood. For pedagogical purposes we assume that the black hole undergoes the following four phases:
\begin{enumerate}[i)]
\item \textit{Balding phase}. When the black hole forms, it will be a highly asymmetric object with gauge field hair. In the initial stage of the evolution, the black hole hair is shed (mainly by the Schwinger pair production mechanism) and asymmetries are lost via gravitational radiation.
\item \textit{Spin-down phase}. At the end of the balding phase, the highly spinning, neutral black hole loses mass and angular momentum through Hawking and Unruh-Starobinskii radiation.
\item \textit{Schwarzschild phase}. At the end of the spin-down phase, the resulting spherically symmetric black hole continues to evaporate but now in a spherical manner. This results in the gradual decrease of its mass
and the increase of its temperature.
\item \textit{Planck phase}. When the mass and/or the Hawking temperature approaches the fundamental scale $T\sim M\sim \Mf$, the black hole can no be longer described semi-classically. A theory of quantum gravity is necessary to study this phase in detail.
\end{enumerate}
The Schwarzschild phase is by far the simplest to study and the majority of the existing literature is devoted to it (see, for example, Ref.~\refcite{Harris:2003eg}). The black hole decays by emitting energy and particles at a temperature
\begin{equation}
T=\frac{\hbar c}{k_B}\frac{(k+1)}{4\pi r_g}.
\end{equation}
For $r_g\sim 10^{-19} \ \mathrm{m}$ the above formula leads to values
in the range $T\sim 77$ GeV for $k=1$ to $T\sim 629$ GeV for $k=7$ (see Ref. \refcite{Kan08}).
By integrating the Stefan-Boltzmann law $dM/dt\sim T^4$, one obtains a decay time of the order of
\begin{equation}
t\approx\frac{\hbar}{M_\star c^2}\left(\frac{M}{M_\star}\right)^{\frac{k+3}{k+1}}\sim 10^{-26}\ \mathrm{s}.
\end{equation}
Unlike the case of a perfect black body, the gravitational potential surrounding the black hole will partially backscatter matter field modes, with consequent depletion of the outgoing flux
 \begin{equation}
\frac{dE^{(s)}(\omega)}{dt}=\sum_j\int\frac{g_{j, k}^{(s)}(\omega)}{\exp{\omega/T}\pm 1}\frac{d^{k+3}\mathrm{p}}{(2\pi)^{k+3}},
\end{equation}
where $|\mathrm{p}|^2=\omega^2-m^2$. The reflection back down the hole is governed by the coefficient $g_{j, k}^{(s)}(\omega)$, called the grey body factor, while $s$ is the spin of the emitted particle and $j$ its angular momentum quantum number. Because this backscattering is a function of $\omega$, the spectrum is no longer exactly Planckian.

Standard model particles (scalars, fermions and gauge bosons) are emitted only on the brane, while in the bulk there is just emission of gravitational radiation (mostly in the form of gravitons, but potentially also scalars).  The proportion of the total emission which is in the form of unobservable bulk gravitational radiation is an important quantity to understand, as it will contribute to missing energy in the detector.  The number of bulk gravitational degrees of freedom increases rapidly as the number of extra dimensions increases, so that, although there are many degrees of freedom in the standard model particles, for large enough $k$ a significant proportion of the total energy is lost into the bulk\cite{Kan08}.
It can be seen from Tab. \ref{tab1} that, even for a single scalar degree of freedom, in the Schwarzschild phase roughly the same amount of energy is emitted on the brane as in the bulk for seven extra dimensions.
The spin-down phase, during which the black hole is rotating, is more complex but has also been studied in detail,\cite{spindown} although there are only partial results for graviton emission.

In modelling the black hole evaporation, we study continuous emission from a fixed background black hole geometry.
In practice the process is stochastic in nature,\cite{Frost:2009cf} with the continuous
emission results providing probabilities for the emission of a particular type of particle in a particular direction and with a particular energy.
In simulations of black hole events,\cite{Frost:2009cf} the emission proceeds in steps, with the black hole emitting a particle, then settling down into an equilibrium state before the emission of the next particle.

In a realistic particle detector, the situation is even more complex.
Radiated particles can be so energetic as to trigger pair production and bremsstrahlung mechanisms.
As a result the colliding partons are followed by a multiplicity of particles in  the case of  black hole formation, i.e.
\begin{equation}
\mathrm{p} + \mathrm{p} \Rightarrow \mathrm{black}\ \mathrm{hole} \rightarrow \mathrm{``particles''}.
\end{equation}
Both QED and QCD drive pair production of $e^\pm$, $q$, $\bar{q}$ and bremsstrahlung with emission of photons $\gamma$ and gluons $g$ respectively. An electron-positron-photon plasma and a quark-gluon plasma form around the black hole in regions called the photosphere and chromosphere respectively. These occur at different critical temperatures, namely $T_{\mathrm{c}}^{\mathrm{QED}}\sim 50 \ \mathrm{GeV}$ and $T_{\mathrm{c}}^{\mathrm{QCD}}\sim 175 \ \mathrm{MeV}$ respectively. The difference between these temperatures explains why the actual black hole emission is dominated by hadrons, which result from parton fragmentation. Specifically,  one can estimate that for $k>2$, the secondary emission consists of $60 \%$ quarks, $15 \%$ gluons, $10 \%$ leptons, $6 \%$ weak bosons, $5 \%$ neutrinos, $1 \%$ photons and $1 \%$ Higgs bosons.\cite{CaS06}
Due to energy conservation, the proliferation of particles leads to a lower average energy per particle. In other words, the direct Hawking spectrum is turned into an effective black body spectrum with a temperature lower than the black hole temperature. In conclusion, black hole formation can be experimentally recognized as an event with a reduced visible energy (gravitational degrees of freedom emitted in the bulk cannot be detected), a high multiplicity, a hadronic to leptonic activity of about $5:1$ and a highly spherical multi-jet emission.\cite{Cav03}
\begin{table}
\tbl{\label{tab1}The Schwarzschild phase. Ratio of bulk/brane scalar field emission, against the number of extra
dimensions $k$, for Schwarzschild black holes (top line) 
and NCBHs (bottom line).
}
{\begin{tabular}{@{}cccccccc@{}} \toprule
$k$ &  1  & 2  & 3  & 4  & 5  & 6  & 7 \\
 \colrule
Schwarzschild  & 0.40  & 0.24   & 0.22  & 0.24  & 0.33  & 0.52  & 0.93  \\
NCBH & 0.265  & 0.082   & 0.027  & $8.9\times 10^{-3}$  & $2.9\times 10^{-3}$  & $9.5\times 10^{-4}$  & $2.8\times 10^{-4}$ \\
\botrule
\end{tabular}}
\end{table}

\section{The case of quantum gravity improved black holes}

Additional experimental signatures of black hole formation depend on the unknown nature of the Planck phase. There exist two major scenarios. The Schwarzschild phase could end up in a phase characterized by the non-thermal emission of a few hard visible quanta. Alternatively, the evaporation could stop due to the formation of stable zero temperature black hole remnants. Charged remnants would be directly observed in ionizing tracks in detectors, while neutral ones could be detected only through a modified distribution of transverse momentum.\cite{KBH05}
%
%

\begin{figure}
\centerline{\psfig{file=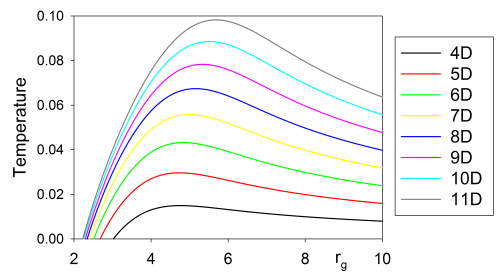,width=6.5cm}}
\caption{\label{fig2}Temperatures of NCBHs in units in which $\ell=1$.
}
\end{figure}

The formation of remnants is supported by recent proposals modeling the Planck phase by means of effective theories of quantum gravity. Due to our inability to describe the last stage of the evaporation by means of full formulations of quantum gravity, one can think about implementing a specific feature we expect from any theory of quantum gravity in the gravitational field equations. One such feature is the emergence of a fundamental length $\ell$, beyond which no further resolution of the space-time manifold is possible\cite{Hossenfelder:2003jz}. 
This line of reasoning has led to the derivation of some families of quantum gravity improved black hole space-times (QGBHs).\cite{BoR00}
In spite of the different derivations, these new effective quantum geometries tend to agree on a common behavior: the curvature singularity is cured and there is the possibility of horizon extremisation even for the neutral, non-rotating case. This implies, on the thermodynamic side, the presence of a phase transition to a positive heat capacity cooling down phase (see Fig. \ref{fig2}).
QGBHs are colder compared to their classical analogues, a fact that permits a semi-classical description of the evaporation without a breakdown of the formalism. Eventually the black hole evaporation switches off, with subsequent remnant formation.

\begin{figure}
\centerline{\psfig{file=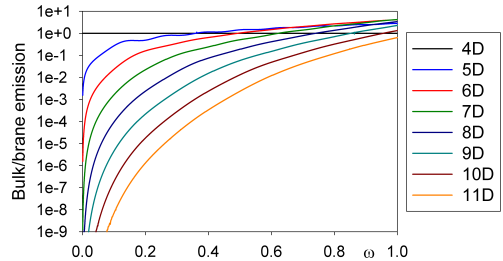,width=6.6cm}}
\caption{\label{fig3}
Bulk/brane emission ratios for scalar field radiation from NCBHs, as a function
of frequency $\omega $ of the emitted quanta, in units in which $\ell =1$.
}
\end{figure}

There are also repercussions on the phases preceding the Planck phase. Recent studies based on one of the families of QGBHs, namely noncommutative geometry inspired black holes (NCBHs),\cite{Nic05}
showed that
emission in the bulk is highly suppressed.
By increasing $k$, the energy emitted in the bulk relatively decreases with respect that emitted on the brane (see Tab. \ref{tab1} and Fig. \ref{fig3})\cite{NiW11}.
In addition NCBHs live significantly longer, having an estimated decay time that does not exceed $10^{-16}\ \mathrm{s}$ irrespective of the kind of brane or bulk emission.\cite{CaN08}
As a price to pay, remnant masses, i.e. the minimum mass for black hole formation, would exceed the energy accessible to the LHC, being of the order of $16$ TeV for $k=1$ and increasing rapidly as $k$ increases.\cite{NiW11} If no other effects come into play, the only chance to observe these objects would be in ultra high energy cosmic ray showers hitting higher layers of the Earth's atmosphere.


\section{Conclusions}

The quest for signatures of the formation of microscopic black holes in particle detectors or cosmic rays is a fascinating and ambitious program. In this short paper we showed how their observation is connected to deeper problems concerning the very nature of quantum gravity. We have presented both an overview of the state of the art in this field and the latest findings based on quantum gravity improved black hole metrics. We stress that the work in this field is ongoing and many questions are yet to be settled. Apart from the issue of drawing a robust and widely accepted scenario for the evaporation end point, there exists a list of open problems concerning the initial phases. For brevity we only recall that the mechanism of the collapse, the computation of the production cross section, the duration of the balding and the spin-down phases, the role of color fields, and the effects of the brane tension are currently subjects of investigations. We stress that these issues are very topical: both the LHC and the Pierre Auger Observatory are now collecting data and soon consistent explanations of experimental results will be necessary.

\section*{Acknowledgements}

This work is supported by the Helmholtz International Center for FAIR within the
framework of the LOEWE program (Landesoffensive zur Entwicklung Wissenschaftlich-\"{O}konomischer Exzellenz) launched by the State of Hesse.  This work is also supported in part by the European Cooperation in Science and Technology (COST) action MP0905 ``Black Holes in a Violent Universe''.
The work of EW is supported by STFC (UK), grant number ST/J000418/1.


\end{document}